\documentclass[twocolumn]{paper} 

\renewcommand{\d}{\mathrm{d}}
\newlength{\fsize}
\begin{document}


\title{High-redshift quasars lensed by spiral galaxies}
\author{Matthias Bartelmann\\ 
        \small 
        Max-Planck-Institut f\"ur Astrophysik, P.O.~Box 1523, 
        D--85740 Garching, Germany} 


\date{\em submitted to Astronomy \& Astrophysics}


\begin{abstract}
Given its extraordinary spatial resolution and sensitivity, the
projected Next Generation Space Telescope (NGST) is likely to detect a
large number of high-redshift QSOs lensed by spiral galaxies. Using
realistic models for the QSO and spiral populations, we calculate the
expected number density of detectable QSOs multiply imaged by spirals,
and investigate the influence of various evolution effects on that
number density. It is shown that NGST will probably find of order ten
lensed QSOs per square degree at 26th magnitude in the V and L bands,
and that various observable quantities like the total number density
of lensed QSOs in these two bands, the ratio between the number
densities of lensed QSOs in the V and L bands, the fraction of QSOs
with more than two images and so forth can be used to constrain the
evolution of the QSOs, the spirals, the dust in spirals, and the
masses of spiral disks.


\end{abstract}

\maketitle 

\section{Introduction}

While strong lensing of QSOs by elliptical galaxies is a well-evolved
subject (e.g.~Turner, Ostriker \& Gott 1984; Fukugita \& Turner 1991;
Maoz \& Rix 1993; Kochanek 1996a,b; Keeton, Kochanek \& Seljak 1997;
Falco, Kochanek \& Mu\~noz 1998; Keeton, Kochanek \& Falco 1998),
interest in lensing by spiral galaxies has only started to develop
over the past few years (Maller, Flores \& Primack 1997; Keeton \&
Kochanek 1998; Bartelmann \& Loeb 1998; Blain, M{\o}ller \& Maller
1999). This was legitimate for the interpretation of QSO surveys
undertaken at angular resolutions of order $1''$. Once angular
resolutions on the order of $0.1''$ can be achieved, a substantial
portion of the lens population may be spiral galaxies. If
observations at high angular resolution can be combined with high
sensitivity, numerous QSOs multiply imaged by spiral galaxies can be
expected to be observed, and such observations can provide detailed
information on the evolution of spiral galaxies and their properties
out to fairly high redshift.

This paper addresses the question of how many QSOs lensed by spiral
galaxies can be expected to be observed by telescopes like NGST, whose
angular resolution exceeds $0.1''$, and which reach a sensitivity of
order one nano-Jansky in the visible and near-infrared wave
bands. Going to such observational limits, QSOs lensed by spiral
galaxies should be discovered in substantial numbers. Many of the
multiple-image systems produced by spiral galaxies will straddle
near-edge-on galactic disks. The statistics of image separations and
time delays will therefore allow to determine properties of spiral
disks at moderate and high redshifts. The wavelength regime between
the visible and the near infrared is particularly interesting for such
investigations, since the importance of dust changes dramatically
across this wavelength range. While multiple QSO images, in particular
near galactic disks, are substantially obscured by dust in the V band,
they are almost unaffected in the L band. Comparing observations taken
in these two wavebands, it will therefore be possible to quantify the
amount and the distribution of dust in distant spiral disks.

The very faint flux limits envisaged will bring high-redshift QSOs
into view. The QSO population at high redshifts is
uncertain. Observations show a quick drop of the QSO number density
beyond redshifts of $4-5$ (Pei 1995). This may reflect rapid QSO
evolution at these redshifts, but also incompleteness of the QSO
samples observed. We therefore investigate two alternatives, assuming
either that the observed QSO population is complete, or that it is to
be extrapolated towards high redshifts.

It is also important for this study to reliably estimate QSO
$K$-corrections in the visible and the near infrared. For this
purpose, we model the dominant features of QSO spectra together with
the Lyman-$\alpha$ forest, and determine $K$-corrections by convolving
the redshifted model spectrum with the appropriate filter functions.

Finally, the contribution of disks to lensing by spiral galaxies is
crucial for two reasons. First, projected disks at large inclination
angles reach high surface mass densities and affect multiple-imaging
cross sections substantially. Second, dust in these disks demagnifies
images and thereby alters the statistics of observed image properties.

We proceed in the following way. Section~2 describes the computation
of QSO $K$-corrections and summarises the model for the QSO luminosity
function. Our model for the spiral galaxies is introduced in
Section~3. Section~4 briefly describes the numerical method. Section~5
presents the results, and we finish with the conclusions in Section~6.

Throughout, we adopt a CDM cosmological model with $\Omega=0.3$,
$\Omega_\Lambda=0$, and a Hubble constant of $H_0=70\,{\rm
km\,s^{-1}\,Mpc^{-1}}$. The CDM spectrum is normalised such that the
local abundance of rich galaxy clusters is reproduced (Eke, Cole \&
Frenk 1996; Viana \& Liddle 1996).

\section{QSOs}

\subsection{QSO spectra and the $K$-correction}

A reliable estimate for the QSO $K$-correction is very important for
the present study. We therefore have to construct a synthetic QSO
spectrum representing the prominent features of a typical QSO
spectrum. This spectrum consists of the underlying QSO continuum, the
hydrogen emission lines of the Lyman, Balmer, and Paschen series, the
corresponding Balmer and Paschen continua, the two-photon continuum,
the $3000\,$\AA\ bump mainly contributed by iron (Fe~II) emission
lines, and emission lines of several other elements like helium,
carbon, nitrogen, oxygen and so forth.

\begin{figure}[ht]
  \centerline{\includegraphics[width=\fsize]{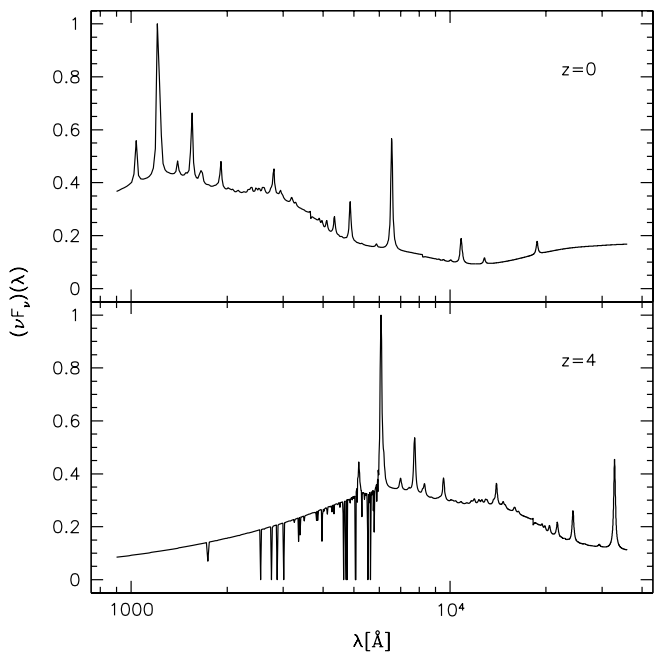}}
\caption{Synthetic QSO spectra at redshifts $z=0$ and $z=4$ (top and
bottom panel, respectively), including strong Lyman-$\alpha$ forest
lines. Wavelengths range from $900\,$\AA\ to $36,000\,$\AA. The
ordinate is scaled to arbitrary units. The most prominent line in both
spectra is Lyman-$\alpha$.}
\label{fig:q1}
\end{figure}

We adopt the QSO continuum emission from Elvis et al.~(1994), who
tabulate the median QSO continuum in the frequency range
$10^9\,\mathrm{Hz}\le\nu\le10^{19}\,\mathrm{Hz}$. The spectrum on top
of the broad continuum is modelled as in Grandi (1981, 1982). It
contains the Balmer- and Paschen series up to order 28 with line
strengths adopted from Brocklehurst (1971), exponential Balmer- and
Paschen continua, the two-photon continuum as given by Spitzer \&
Greenstein (1951), 79 emission lines of Fe~II to reproduce the
$3000\,$\AA\ bump, and several other emission lines of helium, carbon,
nitrogen, oxygen, magnesium, and silicon. The electron temperature is
assumed to be $12,500\,K$, the electron density is
$10^6\,\mathrm{cm}^{-3}$, and the lines are broadened by Lorentzian
profiles with an FWHM velocity of $4,500\,\mathrm{km\,s}^{-1}$. Some
additional lines tabulated by Netzer \& Davidson (1971) are
added. Following Grandi (1982), the emission lines are normalised such
that the equivalent width of the H$\beta$ line against the continuum
is $100\,$\AA. Finally, the number density and equivalent widths of
Lyman-$\alpha$ forest lines are modelled following Murdoch et
al.~(1986). A synthetic spectrum of a QSO at redshift $z=4$ obtained
in this way is displayed in Fig.~(\ref{fig:q1}).

\begin{figure}[ht]
  \centerline{\includegraphics[width=\fsize]{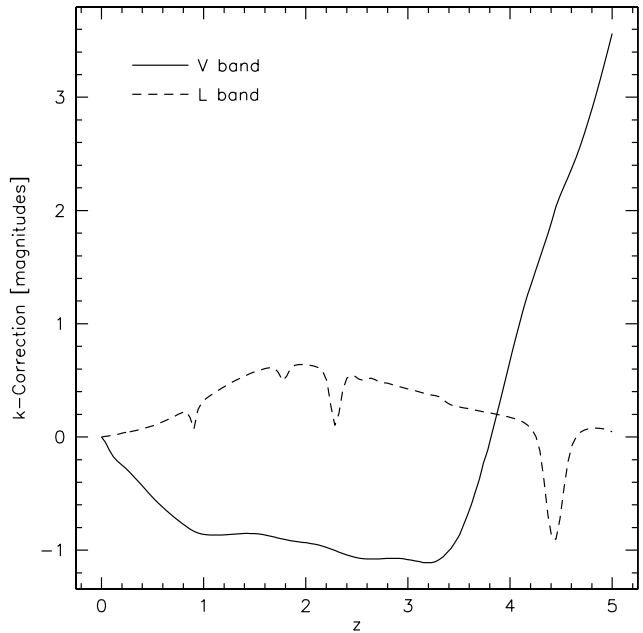}}
\caption{$K$-corrections as functions of QSO redshifts in the V and L
bands (solid and dashed lines, respectively), computed from synthetic
QSO spectra convolved with the V and L response functions.}
\label{fig:q2}
\end{figure}

The synthetic QSO spectrum is then convolved with the response
functions of the Johnson V and L bands to obtain the
$K$-correction. The peak transmission wavelengths of the response
curves are at $0.555\,\mu\mathrm{m}$ for the V band and at
$3.6\,\mu\mathrm{m}$ for the L band. The $K$-corrections in these
bands are plotted in Fig.~(\ref{fig:q2}).

The $K$-correction in the V band is negative from redshift zero to
just above redshift 3, where it rises steeply. Negative
$K$-corrections are due to bright emission lines moving into the
V-band response function, while the steep increase is caused by
Lyman-$\alpha$ forest absorption. In the L band, the $K$-correction
ranges between $\pm1$ magnitudes. The shallow peak in the L-band
$K$-correction is due to the shallow valley in the QSO spectrum beyond
$\approx8,000\,$\AA. The pronounced minimum at $z\approx4.4$ is caused
by the Balmer-$\alpha$ line's coming into the L-band response
function.

\subsection{QSO luminosity function at high redshift}

Haiman \& Loeb (1998) recently provided a model for the QSO luminosity
function at high redshifts. The model rests on four simple and
intuitive assumptions: (i) black holes form in all dark-matter haloes
above a mass limit $M_0$; (ii) the black-hole mass $M_\mathrm{bh}$ is
a certain constant small fraction $\epsilon$ of the halo mass; (iii)
black holes produce QSOs shining at the Eddington luminosity
$L_\mathrm{edd}=L_{\mathrm{edd},\odot}\,M_\mathrm{bh}/M_\odot$ at
birth, and the QSO luminosity is on average a monotonically decreasing
function $M_\mathrm{bh}\,f(t)$ of time $t$ after birth; and (iv) the
mass function of dark-matter haloes is given by the Press-Schechter
(1974) mass function $N(M)$.

The QSO luminosity function is then
\begin{equation}
  \Phi(L,z)=\int_\infty^z\,\d z'\,\int_{M_0}^\infty\,\d M\,
  \frac{\d N(M)}{\d z'}\,\delta[L-M_\mathrm{bh}f(t)]\;.
\label{eq:q1}
\end{equation}
The derivative of the Press-Schechter mass function with respect to
$z$ gives the birth rate of haloes and thus the birth rate of black
holes, and the delta function selects those QSOs shining with
luminosity $L$ at time $t$ after their birth.

The redshift derivative of the Press-Schechter mass function contains
a negative contribution because of halo merging. It therefore
underestimates the halo formation rate. However, the merger
probability at high redshift is low, so that the underestimate is
unimportant (see also Sasaki 1994).

The black-hole mass fraction $\epsilon$ is kept constant here for
simplicity. Introducing a moderate scatter in $\epsilon$
(cf.~Magorrian et al.~1998; Van der Marel 1999) affects the black-hole
mass function only weakly (Haiman \& Loeb 1999).

Haiman \& Loeb used an exponential function for $f(t)$ and assumed
$M_0=10^8\,M_\odot\,[(1+z)/10]^{-3/2}$ for the minimum halo mass. The
model has then only two free parameters, the black-hole mass fraction
$\epsilon$ and the time constant $t_0$ of $f(t)$. Haiman \& Loeb
showed that $\epsilon\approx5.4\times10^{-3}$ and
$t_0\approx6.3\times10^5\,\mathrm{yr}$ provide an excellent fit to the
observed QSO luminosity function at intermediate redshifts (Pei 1995).

Carrying out the redshift integral in (\ref{eq:q1}) leads to
\begin{equation}
  \Phi(L,z)=\int_{M_0}^\infty\,\frac{\d M}{M|\dot{f}(t)|_{t_*}}\,
  \left.\frac{\d N(M)}{\d z}\right|_{z_*}\,
  \left.\frac{\d z}{\d t}\right|_{z_*}\;.
\label{eq:q2}
\end{equation}
Here, $z_*$ is the redshift at which a QSO in a halo of mass $M$ needs
to be born in order to reach luminosity $L$ at redshift $z$, and $t_*$
is the time elapsed between $z_*$ and $z$. Note that another mass
limit on the haloes is implicit here. The QSO cannot shine with higher
than Eddington luminosity. The halo mass therefore needs to exceed
both $M_0$ and $(M_\odot/\epsilon)\,(L/L_{\mathrm{edd},\odot})$.

We repeat the derivation of Haiman \& Loeb's luminosity function here
because numerical evaluation reveals that sufficient approximations
can be made to render the remaining integral in (\ref{eq:q2})
analytically tractable. This is convenient because the accurate
numerical evaluation of (\ref{eq:q2}) requires substantial care
because of the large dynamic range of the Press-Schechter mass
function.

We introduce two approximations. The first is that $t_*$ is typically
very short compared to the Hubble time, so that $z_*\approx z$ is a
very good approximation. The second is that the power spectrum $P(k)$
of the dark-matter density fluctuations can be written as a power law,
\begin{equation}
  P(k)\propto k^n\;.
\label{eq:q3}
\end{equation}
This approximation is well satisfied in the range of scales $k$
corresponding to the relevant dark-matter halo masses, where
$n\approx-2.4$. The first approximation removes the necessity to
numerically evaluate $z_*$. The second approximation allows to write
the Press-Schechter mass function as
\begin{equation}
  N(M)=\frac{\bar{\rho}}{M_*^2}\,\left(\frac{2}{\pi}\right)^{1/2}
  \frac{\alpha}{D(a)}\,m^{\alpha-2}\,
  \exp\left(-\frac{m^{2\alpha}}{D^2(a)}\right)\;,
\label{eq:q4}
\end{equation}
where $\bar{\rho}$ is the mean cosmic density {\em today\/}, $M_*$ is
the nonlinear mass scale {\em today\/}, $m\equiv M/M_*$ is the halo
mass in units of $M_*$, $D(a)$ is the growth factor of density
perturbations as a function of the scale factor $a$, and $\alpha$ is
related to $n$ through
\begin{equation}
  \alpha\equiv\frac{n+3}{6}\;.
\label{eq:q5}
\end{equation}

With (\ref{eq:q4}), the remaining mass integral in (\ref{eq:q2}) can
be carried out. The final result for the luminosity function is
\begin{eqnarray}
  \Phi(L,z)&=&\frac{t_0}{L}\,\frac{\d z}{\d t}\,
  \frac{\bar{\rho}}{M_*}\,
  \frac{2^\beta}{\sqrt{2\pi}}\,a^2D'(a)D^{2\beta-2}(a)
  \nonumber\\
  &\times&
  \left[2\Gamma(1+\beta,\mu_0)-\Gamma(\beta,\mu_0)\right]\;,
\label{eq:q6}
\end{eqnarray}
where we introduced the abbreviations
\begin{equation}
  \beta\equiv\frac{\alpha-1}{2\alpha}
\label{eq:q7}
\end{equation}
and
\begin{equation}
  \mu\equiv\frac{m^{2\alpha}}{2D^2(a)}\;.
\label{eq:q8}
\end{equation}
As mentioned before, $\mu_0$ is determined through (\ref{eq:q8}) by
\begin{equation}
  M_0=\max\left(
    M_0,\frac{LM_\odot}{L_{\mathrm{edd},\odot}\epsilon}
  \right)
\label{eq:q9}
\end{equation}
and $m\equiv M/M_*$. For low QSO luminosities, the term in brackets in
(\ref{eq:q6}) is constant in luminosity, and $\Phi(L,z)\propto
L^{-1}$.

For $n=-2.4$, $\alpha=0.1$ and $\beta=-4.5$. In an Einstein-de Sitter
universe, $D(a)=a$ and $\d z/\d t=H_0\,a^{-5/2}$, and thus
\begin{eqnarray}
  \Phi(L,z)&=&\frac{t_0H_0}{L}\frac{\bar{\rho}}{M_*}
  \frac{2^\beta}{\sqrt{2\pi}}\,a^{2\beta-5/2}\nonumber\\
  &\times&
  \left[2\Gamma(1+\beta,\mu_0)-\Gamma(\beta,\mu_0)\right]\;.
\label{eq:q10}
\end{eqnarray}
Equation~(\ref{eq:q10}) shows that the QSO luminosity function
$\Phi(L,z)$ is a very steep function of redshift.  Ignoring the
redshift dependence of $M_0$, $\Phi(L,z)$ scales with $a$
approximately as $a^{-23/2}$, strongly depending on the local slope of
the dark-matter power spectrum.

\subsection{QSO number counts}

Haiman \& Loeb's model is valid at high redshifts. At low redshifts,
the QSO luminosity function can be assumed to be sufficiently well
observed. In order to test the impact of the theoretical extrapolation
on the expected counts of strongly lensed QSOs, we use two alternative
QSO luminosity functions, namely either Pei's (1995) fit to the
observed QSO number counts alone, or Pei's function augmented by
Haiman \& Loeb's (1998) theoretical model. In the latter case, we
construct the luminosity function $\Phi(L,z)$ from the observed
function out to redshift 3, from the theoretical model beyond redshift
4, and from a redshift-dependent linear combination of the two between
redshifts 3 and 4. For simplicity, we shall refer to these two
luminosity functions as the ``observed'' and the ``theoretically
extrapolated'' one below.

Using the $K$-correction derived in \S~2.1, the QSO luminosity
function can now be converted to number counts per flux and redshift
intervals $\d S$ and $\d z$,
\begin{equation}
  \frac{\d^2N(S,z)}{\d S\d z}\,\d S\d z\;.
\label{eq:q11}
\end{equation}

\section{Spiral galaxies}

\subsection{Evolution of spirals}

We assume that the spiral population has a Schechter (1976) luminosity
function, i.e.~the number density $n$ follows
\begin{equation}
  \frac{\d n(l)}{\d l}=n_*\,l^\nu\,\exp(-l)\;,
\label{eq:s1}
\end{equation}
where $l=L/L_*$ is the luminosity in units of $L_*$,
$n_*=1.5\times10^{-2}\,h^3\,{\rm Mpc}^{-3}$ is the present-day number
density, and $\nu=-0.81$ (Marzke et al.~1994). We further assume that
the circular velocity scales with $l$ according to the Tully-Fisher
(1977) relation,
\begin{equation}
  \frac{v_\mathrm{c}}{v_{\mathrm{c}*}}=l^{1/\alpha}\;,
\label{eq:s2}
\end{equation}
with $\alpha=4$. We adopt two scenarios for the redshift evolution of
the spiral population, namely either constant comoving number density,
or a number density evolving according to the Press-Schechter mass
function.

Mo, Mao \& White (1998) suggested a model for the evolution of spiral
disks resting upon four simple and intuitive assumptions. These are:
(i,ii) The disk mass and angular momentum are fixed fractions of the
spiral mass and angular momentum. (iii) The radial disk profile is
exponential, and the disk is centrifugally supported. (iv) The disk is
dynamically stable.

These assumptions yield a set of simple scaling relations for the disk
properties. Disk mass $M_\mathrm{d}(z)$ and radius $r_\mathrm{d}(z)$
scale with the inverse Hubble parameter $H(z)$,
\begin{equation}
  (M,r)_\mathrm{d}\propto\left[\frac{H(z)}{H_0}\right]^{-1}\;,
\label{eq:s3}
\end{equation}
and consequently the disk surface mass density $\Sigma_\mathrm{d}(z)$
scales with the Hubble parameter,
\begin{equation}
  \Sigma_\mathrm{d}(z)\propto\frac{H(z)}{H_0}\;.
\label{eq:s4}
\end{equation}
We use either constant disk parameters below or evolve them according
to eqs.~(\ref{eq:s3}) and (\ref{eq:s4}).

Finally, we assume that the disk radius $r_\mathrm{d}$ scales with
luminosity as
\begin{equation}
  r_\mathrm{d}\propto l^{1/2}\;.
\label{eq:s5}
\end{equation}

\subsection{The lens model}

Our mass model for the spirals consists of two components, a halo and
a disk. They are modelled as described in Keeton \& Kochanek (1998)
and Bartelmann \& Loeb (1998). We refer the reader to these papers for
details. In essence, both components are described as isothermal,
arbitrarily flattened, ellipsoidal mass distributions, characterised
by a core radius $s$ and an oblateness $q$, defined as the ratio
between the short and the long axes of an oblate ellipsoid. We can
then compose the lensing potential $\psi$ of a spiral with a maximal
disk of three building blocks,
\begin{equation}
  \psi_\mathrm{max}=\psi(s_\mathrm{d},q_\mathrm{d})-
  \psi(r_\mathrm{d},q_\mathrm{d})+\psi(s_\mathrm{h},1)\;.
\label{eq:l1}
\end{equation}
The first two terms are the lensing potential of the disk, truncated
at radius $r_\mathrm{d}$, the third is the lensing potential of the
halo. For sub-maximal disks with a disk fraction $0\le
f_\mathrm{d}\le1$, this is changed to
\begin{equation}
  \psi=f_\mathrm{d}\,\psi_\mathrm{max}+(1-f_\mathrm{d})\,
  \psi(s_\mathrm{h},1)\;,
\label{eq:l2}
\end{equation}
i.e.~the truncated maximal-disk model is scaled down by $f_\mathrm{d}$
and embedded in a spherical halo of unchanged mass. We therefore have
five parameters in total for the mass model, namely the core radius of
the disk $s_\mathrm{d}$, the core radius of the halo $s_\mathrm{h}$,
the oblateness of the disk $q_\mathrm{d}$, the disk fraction
$f_\mathrm{d}$, and the asymptotic circular velocity $v_\mathrm{c}$.

For an $L_*$ galaxy, we assume a circular velocity of
$v_\mathrm{c*}=220\,\mathrm{km\,s}^{-1}$ and a disk radius of
$r_\mathrm{d*}=8\,h^{-1}\,$kpc. They scale with spiral luminosity as
given by eqs.~(\ref{eq:s2}) and (\ref{eq:s5}).

Flat rotation curves require a halo core radius of
$s_\mathrm{h}\approx0.72\,r_\mathrm{d}$ (Keeton \& Kochanek 1998), and
for the disk core radius we assume
$s_\mathrm{d}=0.2\,h^{-1}\,$kpc. The disk fraction is either set to 1
or $0.5$.

Up to a constant, the light travel time to the observer from a source
at angular position $\vec\beta$ through an image position $\vec\theta$
is given by
\begin{equation}
  t(\vec\theta)=\frac{1+z'}{c}\,\frac{D(z')D(z)}{D(z',z)}\,
  \left[\frac{1}{2}(\vec\theta-\vec\beta)^2-\psi(\vec\theta)\right]\,,
\label{eq:l3}
\end{equation}
where $z'$ and $z$ are the lens and source redshifts, respectively;
$D(z',z)$ is the angular-diameter distance between redshifts $z'$ and
$z$, and $D(z)\equiv D(0,z)$. The lens equation relates $\vec\theta$
and $\vec\beta$ through
\begin{equation}
  \vec\beta=\vec\theta-\vec\alpha(\vec\theta)\;,
\label{eq:l4}
\end{equation}
where $\vec\alpha(\vec\theta)=\nabla\psi(\vec\theta)$ is the
deflection angle. The time delay between images is therefore
determined by the lensing potential, its gradient, and the geometry of
the lens system.

\subsection{Gas and dust}

The disk is further assumed to encompass a double-exponential
neutral-hydrogen disk with a scale radius given by $r_\mathrm{d}$, a
scale height of $q_\mathrm{d}r_\mathrm{d}$, and a central column
density of $11.25\times10^{20}\,\mathrm{cm}^{-2}$ (Broeils \& van
Woerden 1994). Dust follows the neutral-hydrogen distribution with a
present-day dust-to-gas ratio of $0.01$ (Whittet 1992). The dust is
assumed to be a mixture of silicates and graphites, for which the
extinction curve as a function of wavelength was derived by Draine \&
Lee (1984). The dust-to-gas ratio is either assumed to be constant
with redshift, or to evolve in proportion to $(1+z)^{-2}$ (Pei, Fall
\& Bechtold 1991).

\subsection{Summary of the spiral model}

Having fixed the disk radius, the disk core radius, the halo core
radius, and the asymptotic circular velocity, and assuming the
scalings of the circular velocity and the disk radius with luminosity,
the mass model for the spirals has two more free parameters, namely
the disk fraction $f_\mathrm{d}$ and the disk oblateness
$q_\mathrm{d}$.

The number density of spirals is then either assumed to be constant or
to evolve according to the redshift evolution of the Press-Schechter
mass function in the appropriate mass range. The parameters of
individual spirals are either assumed to be constant with redshift, or
to evolve according to the model by Mo et al.~(1998). The dust-to-gas
ratio in the spiral disks finally is either assumed to be constant or
to evolve proportional to $(1+z)^{-2}$.

We focus on a fiducial model with disk oblateness $q_\mathrm{d}=0.2$,
disk fraction $f_\mathrm{d}=1$, and evolving disk parameters, number
density, and dust.

\subsection{Visibility of faint QSO images}

Many QSO images lensed by spiral galaxies will appear close to spiral
disks. In order to be detectable, they need to be sufficiently
brighter than the disk itself. We can estimate a maximum QSO magnitude
in the following way. Freeman's (1970) law asserts that the central
surface brightness of nearby spirals is $\approx21.5\,{\rm
mag\,arcsec^{-2}}$ in the B band, corresponding to $I_{0,{\rm
V}}\approx20.8\,{\rm mag\,arcsec^{-2}}$ in the V band. Ignoring
spectral evolution, this changes to
\begin{equation}
  I_{0,{\rm V}}(z)=I_{0,{\rm V}}-2.5\,(\alpha-3)\,\log_{10}(1+z)\;,
\label{eq:l5}
\end{equation}
at higher redshift, where $\alpha\approx-1$ is the power-law exponent
of the spectral-energy distribution. The most probable lens redshift
for high-redshift QSOs is $z\approx0.5-0.6$, where the central surface
brightness is $I_{0,{\rm V}}(0.6)\approx22.6\,{\rm mag\,arcsec^{-2}}$.

Typical image separations from the lens centre are of order the
Einstein radius, or roughly $1\,h^{-1}\,{\rm kpc}$. Assuming a disk
scale length of $\approx8\,h^{-1}\,{\rm kpc}$ and an exponential light
profile, the disk surface brightness near image locations has dropped
by a factor of $\approx0.9$ to $I_{\rm V}(0.6)\approx22.7\,{\rm
mag\,arcsec^{-2}}$. NGST is planned to have a pixel size of order
$5\times10^{-4}\,{\rm arcsec}^2$, hence the per-pixel disk magnitude
is of order $31\,{\rm mag}$. Assuming that a QSO image covers a few
pixels only, it seems realistic to expect that QSO images brighter
than $\approx28-29\,{\rm mag}$ can safely be detected on top of a
spiral disk at redshift $\approx0.5-0.6$. However, many multiple QSO
images will straddle near-edge-on disks, whose surface brightness
drops rapidly away from their symmetry line.

This estimate applies to the V band. Spiral disks are likely to be
brighter in the near infrared, where QSOs are intrinsically
fainter. However, QSO host galaxies are substantially more prominent
in the near infrared and give rise to characteristic lensing features
like arcs and rings (Kochanek et al.~1999), which help in detecting
lensed QSOs against spiral disks, and in distinguishing them from
surface-brightness fluctuations. Confusion of lensed QSO images with
bright spots of stellar appearance like OB associations or other
surface-brightness fluctuations can also be disentangled through
colour information or highly resolved X-ray observations. The order of
magnitude of the previous estimate should therefore also hold in the L
band.

\section{Method of computation}

We wish to compute the number of QSOs above a certain flux limit $S$
lensed by spiral galaxies such that the images have certain properties
$Q$. Let 
\begin{equation}
  \frac{\d^2N(S,z)}{\d S\d z}\d S\d z
\label{eq:m0}
\end{equation}
be the number of unlensed quasars with flux within $\d S$ of $S$ and
redshift within $\d z$ of $z$. Let further $P(\mu,z;Q)$ be the
probability for a quasar at redshift $z$ to be magnified by a factor
$\ge\mu$ and lensed with image properties $Q$. The number of lensed
quasars in the redshift interval $z_1\le z\le z_2$ above the flux
limit $S$ is then
\begin{equation}
  N(S)=\int_0^\infty\d S'\,\int_{z_1}^{z_2}\d z\,
  P\left(\frac{S}{S'},z;Q\right)\,
  \left[\frac{\d^2N(S,z)}{\d S\d z}\right](S',z)\;.
\label{eq:m1}
\end{equation}

The lensing probability $P(\mu,z;Q)$ is conveniently computed in the
cross-section approach. The cross section $\sigma(\mu,z',z,l;Q)$ is
defined as the area on the source sphere at redshift $z$ in which a
source must fall in order to be lensed with image properties $Q$ and
magnified by $\ge\mu$ by an individual galaxy with luminosity $lL_*$
at redshift $z'$. The lensing probability is then
\begin{eqnarray}
  P(\mu,z;Q)&=&\frac{1}{4\pi D^2(z)}\,
  \int_0^z\d z'\int_0^\infty\,d l\,
  (1+z')^3\,n(l,z')\nonumber\\
  &\times&\frac{\d V(z')}{\d z}\,
  \sigma(\mu,z',z,l;Q)\;,
\label{eq:m2}
\end{eqnarray}
where $(1+z')^3\,n(l,z')$ is the number density of galaxies with
luminosity $l$ at redshift $z'$, $[\d V(z')/\d z]\d z'$ is the proper
volume of a spherical shell of width $\d z'$ and radius $z'$, and
$D(z)$ is the (angular-diameter) distance from the observer to the
source plane.

It is important to note that total image magnifications are not only
determined by lensing in our model, but also by dust extinction in the
galactic disks. Total magnifications therefore range within
$0\le\mu<\infty$ rather than $1\le\mu<\infty$, and therefore $S'$ in
(\ref{eq:m1}) ranges within $0\le S'<\infty$.

The scaling relations (\ref{eq:s2}) and (\ref{eq:s5}) assure that the
cross section $\sigma$ scales with the squared Einstein radius. The
Einstein radius in turn scales with the squared circular velocity, and
hence with $l^{1/2}$. Assuming the scaling (\ref{eq:s5}) of the disk
radius, the disk parameters scale with galaxy luminosity $l$ in the
same way as the Einstein radius. This allows to immediately perform
the $l$-integral in (\ref{eq:m2}). The remaining redshift integral
needs to be performed numerically.

We are left with the determination of the cross sections
$\sigma$. Since the galactic disk defines a symmetry plane randomly
inclined with respect to the line-of-sight, this involves averaging
the cross sections over inclination angles. For fixed inclination
angles, cross sections are computed by placing sources on a grid in
the source plane, finding all images of each source, computing their
magnifications and classifying the image properties, and finding the
area in the source plane in which sources are magnified by $\ge\mu$
and imaged with properties $Q$.

Technically, we increase numerical efficiency by placing the sources
on an adaptive grid in the source plane in order to increase numerical
resolution where needed, i.e.~near caustic curves. Rough image
positions are then computed in a first coarse scan in the lens plane,
followed by a fine scan to refine image positions and to determine
accurate image properties. This technique of having adaptive grids in
both the source and the lens planes increases computational speed by
more than an order of magnitude compared to the straightforward
fixed-grid approach. Different levels of grid refinement in the source
plane are statistically accounted for by assigning appropriate
statistical weights to the sources.

Having found the cross sections $\sigma$, we can evaluate the lensing
probability (\ref{eq:m2}). With the quasar number counts taken from
\S~2.2, the number of quasars above flux limit $S$ and imaged with
properties $Q$ is then computed evaluating (\ref{eq:m1}).

\section{Results}

Multiply imaged QSOs can be recognised as such if the angular
separation $\Delta\theta$ of the images exceeds a certain threshold,
and if the flux ratio $r$ between the images does not exceed a certain
dynamic range. We adopt $\Delta\theta\ge0.1''$ and $r\le20$ throughout
this section; i.e., with ``lensed QSOs'' we mean multiply imaged QSOs
whose images satisfy these criteria. In case of three or more images,
we take $\Delta\theta$ and $r$ between the two brightest images.

\begin{figure*}[ht]
  \includegraphics[width=0.32\hsize]{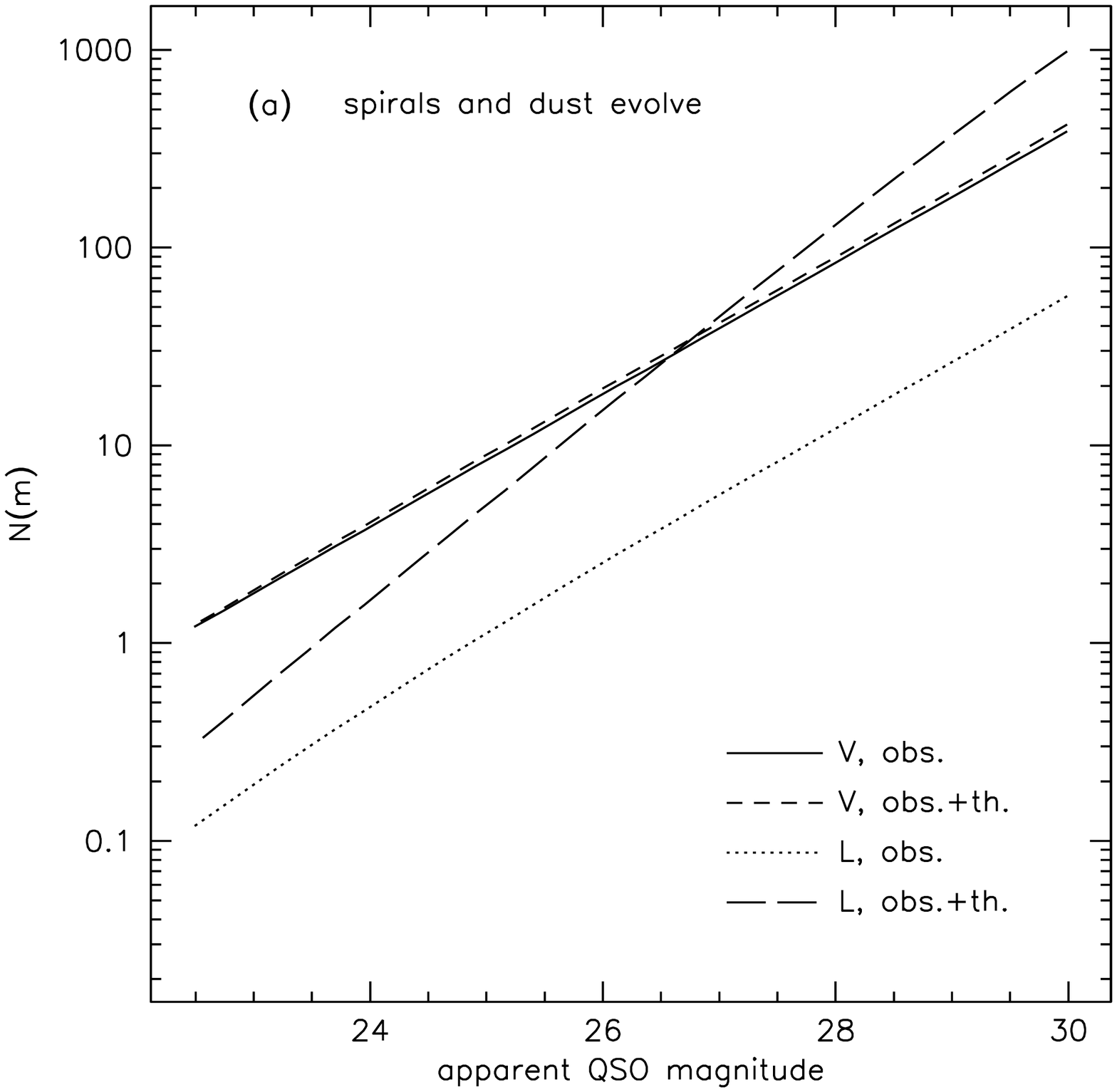}\hfill
  \includegraphics[width=0.32\hsize]{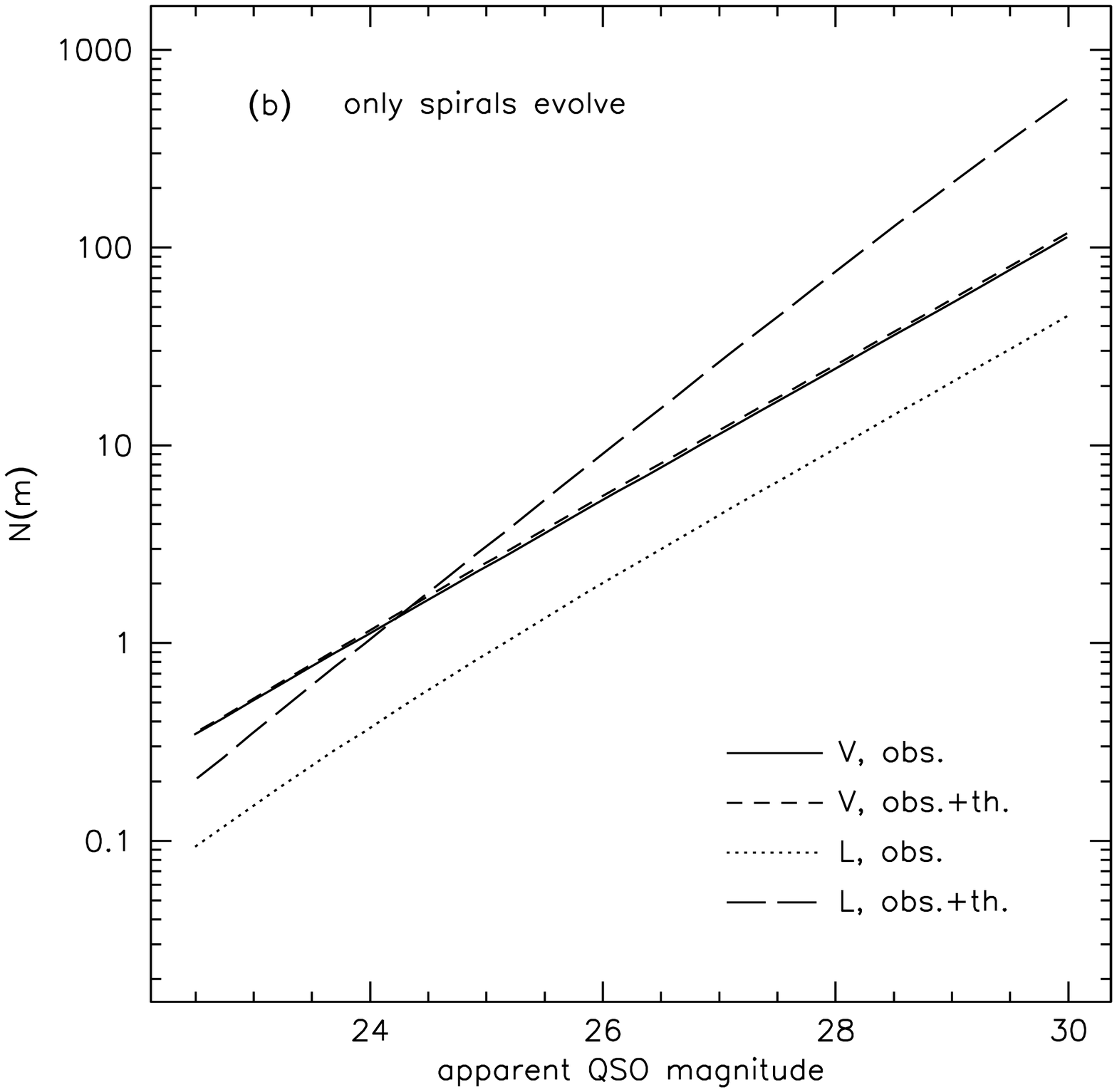}\hfill
  \includegraphics[width=0.32\hsize]{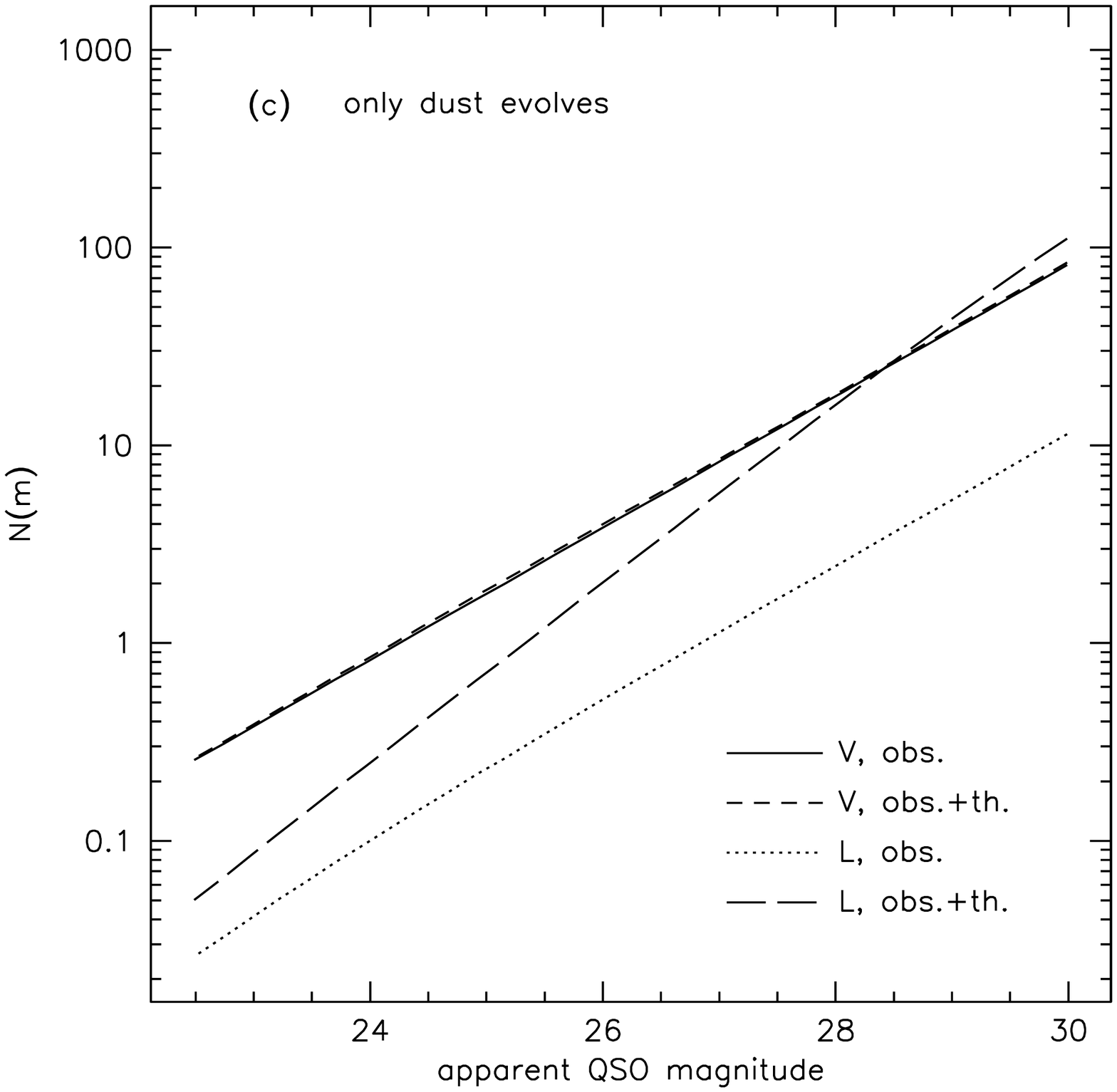}
\caption{Number of quasars on the sky per square degree, lensed by
spiral galaxies such that there are two or more images with image
separation $\ge0.1''$ and flux ratio $\le20$. Each panel has four
curves. Solid and short-dashed curves: V band, QSO luminosity function
fit to the observed number counts only, or extrapolated by the
theoretical model; dotted and long-dashed curve: likewise for the L
band. The three panels show: (a) spiral population and dust evolve;
(b) only spiral population evolves; (c) only dust evolves.}
\label{fig:3}
\end{figure*}

The number density of lensed QSOs per square degree on the sky is
shown in Fig.~\ref{fig:3} as a function of total apparent QSO
brightness. The figure has three panels, and each panel contains four
curves. Panel~(a) shows results for the fiducial model. It assumes a
spiral population with evolving number density and evolving disks,
containing evolving dust. Dust evolution is neglected in panel~(b),
and the evolution of the spiral population is ignored in panel~(c).

The solid and dotted curves show the number density based on the
observed QSO luminosity function in the V and L bands,
respectively. The L-band (dotted) curve falls below the V-band curve
by approximately one order of magnitude because QSOs are intrinsically
fainter in red than in blue light. The remaining two curves (short-
and long-dashed) are again V- and L-band results, now based on the
theoretically extrapolated QSO luminosity function by Haiman \&
Loeb. The changed QSO luminosity function affects the V-band counts
(solid vs.~short-dashed curves) only by a very small amount, while it
has a very significant effect on the L-band results. The reason is the
K-correction. Although the theoretically extrapolated QSO luminosity
function extends to much higher redshifts than the observed one, the
high-redshift QSOs are almost entirely suppressed in the V band
because of the strong K-correction setting in at redshifts
$\gtrsim3.4$. In the L-band, however, the K-correction remains small
or negative even at high redshifts
(cf.~Fig.~\ref{fig:q2}). Consequently, the huge reservoir of
high-redshift QSOs predicted by the theoretically modelled QSO
luminosity function is not at all suppressed in the L-band, where the
different assumptions for the QSO luminosity function result in number
densities differing by approximately one order of magnitude at
apparent magnitudes of $\approx28$. Faintward of about $26.5\,$th
magnitude, the L-band counts exceed the V-band counts.

The neglect of dust evolution in panel~(b) has a strong effect on the
V-band counts, and a substantially smaller effect in the L
band. V-band counts are reduced by factors of $\approx3-4$, regardless
of the assumed QSO luminosity function. L-band counts based on the
observed QSO luminosity function are lowered by $\approx25\%$ compared
to evolving dust, and approximately halved if the theoretically
extrapolated QSO luminosity function is assumed. The reason is of
course that with increasing QSO redshift, the redshift of the most
efficient lenses also increases, so that QSO light in the observer's L
band is more strongly affected by dust in the lenses. This leads to a
relative enhancement of the L-band compared to the V-band counts.

When the spiral number density is assumed to be constant in comoving
space, as in panel~(c), the lensed QSO counts are reduced by almost an
order of magnitude in both bands. This is due to the fact that spirals
merge to form ellipticals towards redshift zero, hence the spiral
number density is expected to increase towards moderate redshifts. If
this increase is neglected, there are fewer lenses available and
therefore fewer lensed QSOs observable.

Variation of other parameters has very little effect on the counts of
lensed QSOs. Lowering the disk mass (by reducing $f_\mathrm{d}$ from
unity to $0.5$), or neglecting the expected change in spiral
properties with redshift, does not change the lensed QSO counts
significantly.

\begin{figure}[ht]
  \centerline{\includegraphics[width=\fsize]{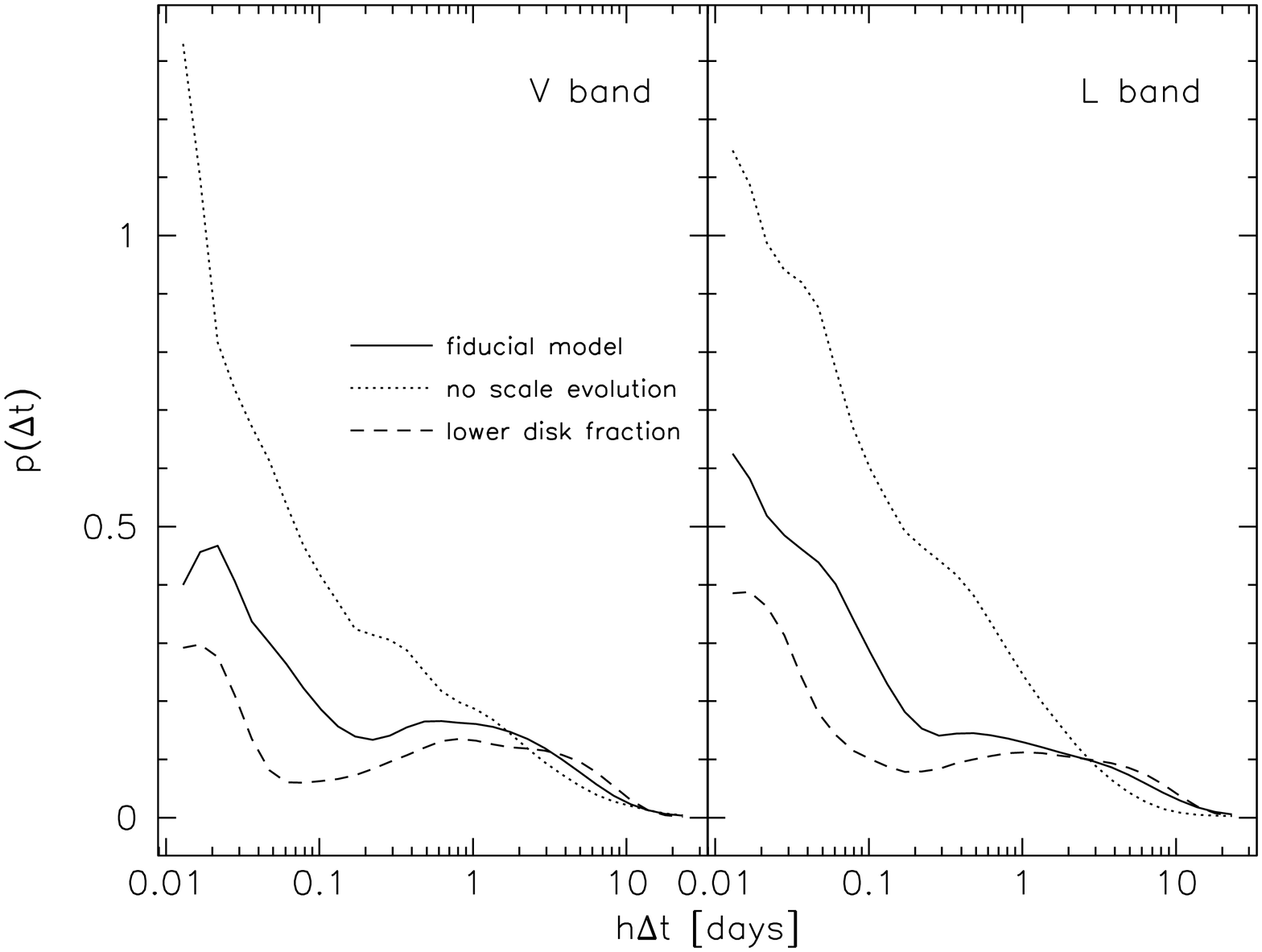}}
\caption{Time-delay distribution for images split by $\ge0.1''$ with
flux ratio $\le20$ and total apparent magnitude $\le25$ in the V and L
bands. Three curves are shown per panel. Solid curve: fiducial model;
dotted curve: no scale evolution; dashed curve: disk fraction reduced
to $f_\mathrm{d}=0.5$.}
\label{fig:4}
\end{figure}

Figure~\ref{fig:4} shows the distribution of time delays between the
two brightest images of lensed QSOs. The figure has two panels for the
two filter bands, as indicated. Each panel shows three curves. The
solid curve is the distribution for the fiducial model, in which
spiral evolution is taken into account, and the disks are maximal. The
curve peaks at time delays of order $0.02$ days, or approximately half
an hour. It goes through a local minimum at $\approx0.2$ days, and
rises again for a broad and shallow peak centred on $\approx1$
day. This bimodal distribution reflects the two components of the mass
distribution. Time delays of order one day occur between images whose
splitting is dominated by the halo, while images straddling a
near-edge-on disk have substantially smaller time delays. This is
further demonstrated by changing the disk contribution to the lens
model. When disk evolution is neglected, as for the dotted curve, the
time-delay distribution peaks more sharply at small time delays, and
the secondary maximum indicative of the halo disappears. Conversely,
the secondary maximum is emphasised relative to the maximum at small
time delays when the disk fraction is reduced to $f_\mathrm{d}=0.5$,
as for the dashed curve.

The secondary maximum in the time-delay distribution is more
pronounced in the V than in the L band. The reason is the dust in the
galactic disks. Images straddling near-edge-on disks, which are
responsible for the peak at low time delays, are strongly suppressed
by dust in the V band, but easily seen in the L band. The primary
maximum in the time-delay distribution is therefore more pronounced in
the L than in the V band, so that the secondary maximum is relatively
lowered.

The results in Fig.~\ref{fig:4} are based on the theoretically
extrapolated QSO luminosity function, but this assumption is
insignificant in this context. Adopting the observed luminosity
function leads to virtually identical results.

\begin{figure}[ht]
  \centerline{\includegraphics[width=\fsize]{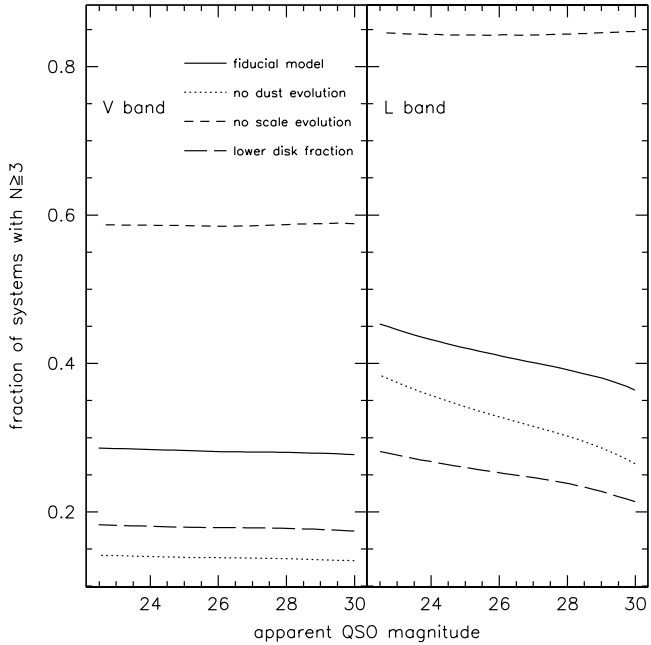}}
\caption{Fraction of multiple image systems with three or more
observable images for the V and L bands. Solid curve: fiducial model;
dotted curve: no dust evolution; short-dashed curve: no scale
evolution; long-dashed curve: disk fraction reduced to
$f_\mathrm{d}=0.5$.}
\label{fig:5}
\end{figure}

Figure~\ref{fig:5} shows the fraction of multiple-image systems with
three or more visible images. As before, images are considered visible
if they are separated by $\ge0.1''$ and span a dynamic range of
$\le20$. Although these systems include such with four or more images,
we refer to the fraction as ``triple fraction'' for simplicity. The
figure also has two panels for the two filter bands considered.

Both panels have four curves. The solid curve represents the fiducial
model, which assumes evolution of the spirals and of the dust in the
spiral disks. There are $\approx30\%$ image systems with more than two
images in the V band, and $\approx40\%$ in the L band. It is again
straightforward that there should be more such systems in the L rather
than in the V band. Dust in the disks suppresses images and therefore
converts part of the triple systems to doubles, and this effect is
much more important in the V than in the L band.

Neglecting dust evolution leads to the dotted curves. The triple
fraction is reduced to just below $10\%$ in the V band and to
$\approx30\%$ in the L band. Obviously, the reduction compared to the
model with evolving dust is more severe in the V than in the L band.

The long-dashed curves are obtained after reducing the disk fraction
from unity to $f_\mathrm{d}=0.5$. Generally, the triple fraction is
reduced, but now the reduction is more severe in the L than in the V
band. The reason for this is that dust suppresses triple systems in
the V band much more strongly than in the L band. Therefore, the
reservoir of triple images in the L band is much larger, so the
relative effect of a lower disk fraction is stronger in the L band.

The strongest effect, however, is achieved by neglecting the evolution
of spiral disks, as shown by the short-dashed curve in
Fig.~\ref{fig:5}. Neglecting this aspect of the evolution leads to
more extended and dominant disks at moderate redshifts. Such disks
cause pronounced caustic curves with narrow, extended spikes following
the projected disks. These extended caustics increase the
multiple-image cross section of spirals considerably. As
Fig.~\ref{fig:5} shows, the triple fraction is increased to just below
$60\%$ in the V band, and to almost $85\%$ in the L band.

\begin{figure}[ht]
  \centerline{\includegraphics[width=\fsize]{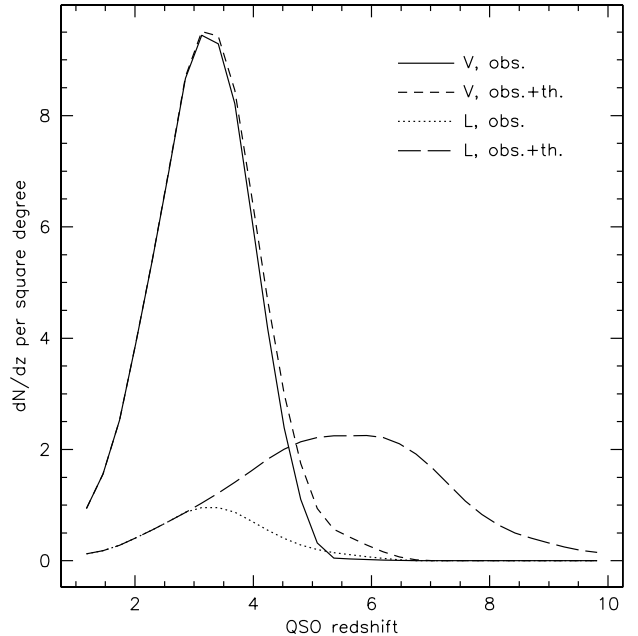}}
\caption{Number $\d N/\d z$ of lensed QSOs with total apparent
magnitude $\le25$ per square degree and unit redshift. Solid curve: V
band, observed QSO luminosity function; dotted curve: L band, observed
QSO luminosity function; short-dashed and long-dashed curves:
theoretically extrapolated QSO luminosity function, V and L band,
respectively.}
\label{fig:6}
\end{figure}

The results shown in Fig.~\ref{fig:5} are based on the theoretically
extrapolated QSO luminosity function. Generally, the V-band results
are unaffected by the choice of luminosity function, and the L-band
results are generally slightly higher if the observed QSO luminosity
function is used instead. The reason is highlighted by
Fig.~\ref{fig:6}.

Figure~\ref{fig:6} shows the redshift distribution of lensed QSOs in
the V and L bands which are brighter than $25\,$th apparent
magnitude. The solid and the dotted curves show the V-band
distribution with the observed and the theoretically extrapolated QSO
luminosity function, respectively. The two distributions are very
similar, and they peak just above $z=3$. The distribution in the L
band based on the observed QSO luminosity function (long-dashed curve)
shows essentially the same behaviour, except that its amplitude is
lower because of the lower L-band QSO counts at $25\,$th
magnitude. Assuming the theoretically extrapolated QSO luminisity
function, however, broadens the peak and shifts it to
$z\approx5-6$. The substantially higher QSO redshifts in the L band
lead to more important dust extinction, and therefore to lower triple
fractions, if the theoretically extrapolated QSO luminosity function
is assumed.

In addition, this explains the different trends with apparent QSO
magnitude of the curves in Fig.~\ref{fig:5}. While the triple fraction
is essentially constant across the plotted range of magnitudes in the
V band, it decreases for fainter QSOs in the L band. This is because
the faint QSO population originates from a broader redshift range in
the L than in the V band, and the importance of dust extinction
increases with QSO redshift.

\section{Conclusions}

We modelled strong lensing of high-redshift QSOs by spiral galaxies,
aiming at future surveys with high angular resolution and high
sensitivity in a wavelength range between the visible and the near
infrared. The main question was, how many multiply imaged QSOs are
expected to be seen, and how do their number and their image
configurations depend on the properties of the lensing spiral
population?

Our mass model for the spiral galaxies consists of a spherical halo
and a disk, which also contains gas and dust. The number density of
the spiral population, and the properties of spiral disks and the dust
therein, are assumed to be either constant in time, or to evolve
according to evolutionary models taken from the literature.

We take the observed QSO luminosity function as parameterised by Pei
(1995) either as it is, or theoretically extrapolated towards high
redshifts (Haiman \& Loeb 1998).

We can summarise our results as follows:

\begin{enumerate}

\item At $28\,$th apparent magnitude, the number density of QSOs
lensed by spiral galaxies reaches almost $100$ per square degree in
the V band, irrespective of whether the observed or the theoretically
extrapolated QSO luminosity function is assumed. The number density of
lensed QSOs per square degree in the L band is just above $10$ for the
observed QSO luminosity function, and well above $100$ for the
theoretically extrapolated luminosity function.

\item Neglecting dust evolution reduces the number density of lensed
QSOs brighter than $28\,$th magnitude by a factor of $\approx3$ in the
V band, and by a factor of $\approx1.5$ in the L band.

\item Neglecting the cosmic evolution of the spiral population reduces
the number density of lensed QSOs brighter than $28\,$th magnitude by
a factor of $\approx5$ in the V and the L bands alike.

\item The time-delay distribution peaks at small time delays below one
hour. It is bimodal with a secondary maximum around one day. The short
time delays are due to images straddling spiral disks, the long time
delays are due to image splitting by the halo. Consequently, the
secondary maximum becomes shallower or higher when the disks become
more or less dominant, respectively.

\item The fraction of image systems with three or more images is of
order $30\%$ in the V band and $40\%$ in the L band for evolving,
maximal disks. It decreases when the disk fraction is reduced, and
increases when the disk evolution is neglected.

\item The redshift distribution of the lensed QSOs peaks around
$z\approx3$ in the V band, regardless of the QSO luminosity function
assumed. The QSO redshift distribution in the L band also peaks around
$z=3$ if the observed QSO luminosity function is assumed, but the peak
broadens and shifts towards $z=5-6$ for the theoretically extrapolated
QSO luminosity function.

\end{enumerate}

These results demonstrate that lensing of high-redshift QSOs by spiral
galaxies will provide a wealth of information once an angular
resolution of $\lesssim0.1''$ can be achieved between the visible and
the near infrared at flux limits of $\lesssim28-30\,$th magnitude.
The projected NGST can reach such flux limits within at most a few
minutes of exposure time. The total number of lensed QSOs per square
degree reflects the number density evolution of spiral galaxies out to
redshifts of $0.5-0.6$. The ratio of the lensed QSO number counts in
the V and L bands is a measure for the dust evolution in the spiral
disks. The time-delay distribution and the fraction of image systems
with three or more images are sensitive to the disk mass and the scale
evolution of the spiral disks.

The disks in spirals are of particular interest here. Spiral haloes
alone have substantially lower lensing capabilities than ellipticals
because of their lower velocity dispersions. A substantial fraction of
the strong-lensing cross sections of spirals is contributed by
near-edge-on disks. If they are sufficiently massive, they give rise
to strongly elongated caustic curves with naked cusps. Lenses with
such caustics produce characteristic image configurations, namely
triple images of comparable brightness with very small angular
separations straddling the disks. The occurrence and abundance of such
image configurations typical for lensing by projected disks will
therefore be a direct measure for the disk fraction in spirals.

\section*{Acknowledgements} 
It is a pleasure to thank Chris Kochanek, Avi Loeb, Paul Schechter and
Peter Schneider for valuable discussions. This work was supported in
part by the Sonderforschungsbereich 375 of the Deutsche
Forschungsgemeinschaft.

\end{document}